\documentclass[aps,pre,showpacs,raggedbottom,nobalancelastpage,amssymb,twocolumn,groupedaddress]{revtex4}
\usepackage{graphicx}
\usepackage{amsmath}
\usepackage{amsfonts}
\usepackage{amssymb}
\usepackage{epsfig}
\usepackage{color}
\usepackage{dsfont}

\newcommand{\abs}[1]{\left\vert#1\right\vert}

\newcommand{\Tr}[1]{\text{Tr}\left\{#1\right\}}
\newcommand{\ParTr}[2]{\text{Tr}_{#1}\left\{#2\right\}}
\newcommand{\bra}[1]{\langle#1\vert}
\newcommand{\ket}[1]{\vert#1\rangle}
\newcommand\braket[2]{\langle#1|#2\rangle}

\begin{document}

\title{Work done in a decoherence process}

\author{Gianluca~Francica}
\email{gianluca.francica@gmail.com}
\noaffiliation

\date{\today}

\begin{abstract}
We investigate the  thermodynamics of a quench realized by turning on an interaction generating pure decoherence. We characterize the work probability distribution function, also with a fluctuation relation, a lower bound of the work and by considering some physical examples.
\end{abstract}

\maketitle
\section{Introduction}
The interaction of an open quantum system with is environment can lead to the phenomenon of decoherence~\cite{zurek91,bookBreuer06}. Specifically, the interaction creates correlations between the states of the system and the environment and destroys superposition of these states in the course of time. In the last decades the non-equilibrium thermodynamics has received a great attention~\cite{kosloff13,Vinjanampathy16,bookthermo18} and recently also the thermodynamics of decoherence has been investigated~\cite{Popovic21}.

In the same spirit of Ref.~\cite{Popovic21}, in this paper we examine the thermodynamics of a quench realized by turning on an interaction generating pure decoherence. In this case the work done does not change the internal energy of the system and is completely converted to heat. In particular, we express the work probability distribution function as a convex combination of probability distribution functions of the environment. Thus, we derive a fluctuation relation characterizing the process and a lower bound of the work done. We proceed with the investigation by considering some physical examples.

\section{Thermodynamics of the process}
We consider a system interacting with its environment, such that for times $t>0$ the total Hamiltonian is of the form
\begin{equation}
H = H_S + H_B + H_I
\end{equation}
where $H_S$ and $H_B$ describe the free evolution of the system and the environment, respectively. The interaction is taken to be
\begin{equation}\label{int}
H_I = \sum_n \ket{n}\bra{n}\otimes B_n
\end{equation}
where $\ket{n}$ are the eigenstates of the system Hamiltonian $H_S$ and $B_n$ are arbitrary environment hermitian operators. We recall that, due to this kind of interaction, an initial state $\ket{\Psi(0)}=\sum_n c_n \ket{n}\otimes \ket{\phi}$, where $\ket{\phi}$ is an arbitrary environment state, evolves into the entangled system-environment state $\ket{\Psi(t)}=U_{t,0} \ket{\Psi(0)}=\sum_n c_n \ket{n}\otimes \ket{\phi_n(t)}$, where the time evolution operator is $U_{t,0}=e^{-i H t}$. By considering the reduced system density matrix $\rho_S(t) = \ParTr{B}{\ket{\Psi(t)}\bra{\Psi(t)}}$, we note that, since $\braket{\phi_n(t)}{\phi_n(t)}=1$, the populations are $\bra{n}\rho_S(t)\ket{n}=\abs{c_n}^2$ and are constant in time. Conversely, the coherences are $\bra{n}\rho_S(t)\ket{m}=c_n c_m^* \braket{\phi_m(t)}{\phi_n(t)}$, with $n\neq m$, and typically the overlap $\braket{\phi_m(t)}{\phi_n(t)}$ rapidly decreases in time, generating the phenomenon of pure decoherence. In particular it is useful to define the decoherence function $\Gamma_{nm}(t)$ (for $n\neq m$) such that $\abs{\braket{\phi_n(t)}{\phi_m(t)}}=e^{\Gamma_{nm}(t)}$.

Here, we take in exam the quench $H_0\to H$, where initially the interaction is turned off, i.e. for times $t\leq 0$ the Hamiltonian reads
\begin{equation}
H_0 = H_S + H_B
\end{equation}
and the total system is prepared in the initial equilibrium state
\begin{equation}
\rho(0) = \frac{e^{-\beta H_0}}{Z_0} = \rho_S(0) \otimes \rho_B(0)
\end{equation}
where the partition function is $Z_0=\Tr{e^{-\beta H_0}}$ and the initial reduced states are $\rho_X(0)=e^{-\beta H_X}/Z_X$ with $Z_X=\Tr{e^{-\beta H_X}}$, for $X=S,B$.
The energy difference of the system after a time $t$ is $\langle \Delta E_S\rangle = \Tr{(U_{t,0}^\dagger H_S U_{t,0}-H_S)\rho(0)}$ and, by noting that $[H,H_S]=0$, we obtain $\langle \Delta E_S\rangle = 0$ for any time $t>0$. We note that, also for strong coupling, the internal energy change, as defined in Ref.~\cite{rivas20}, is equal to $\langle \Delta E_S\rangle=0$ (see Appendix). From a thermodynamic point of view, this means that the work done on the system is equal to the heat exchange with its environment, i.e. $\langle w\rangle = \langle Q \rangle$. In particular, the work reads
\begin{equation}
\langle w \rangle = \Tr{(U_{t,0}^\dagger H U_{t,0}-H_0)\rho(0)} = \Tr{H_I \rho(0)}\,.
\end{equation}
It is worth noting that the work is produced suddenly at the instant in which the interaction is turned on. Thus, during the decoherence process, i.e. for $t>0$, no heat is exchanged with the environment. Anyway, we can realize a pure decoherence process generated by a time-dependent Hamiltonian, for instance by considering time-dependent operators $B_n(t)$ (such that $B_n(0)=0$), such that work, and so heat (always equal to the work), are produced during the decoherence process.
By focusing on the interaction of Eq.~\eqref{int}, with constant $B_n$, the work reads
\begin{equation}
\langle w \rangle = \sum_n p^S_n \Tr{B_n \rho_B(0)}\,,
\end{equation}
where we have defined the system populations $p^S_n=\bra{n}\rho_S(0)\ket{n}$. Thus, the work is the weighted mean of the averages of the environment operators $B_n$, calculated with respect to the state $\rho_B(0)$, with weights $p^S_n$.

By considering a two-measurement scheme, the work can be expressed as $\langle w \rangle = \int w p(w) dw$, where $p(w)$ is the work probability distribution function (see, e.g., Ref.~\cite{campisi11} for its general definition).
By taking in account the eigenvalue equation
\begin{equation}
(H_B+B_n)\ket{E^n_k} = E^n_k \ket{E^n_k}\,,
\end{equation}
the work probability distribution function can be expressed as the convex combination
\begin{equation}\label{prob}
p(w)= \sum_n p^S_n p_n(w)
\end{equation}
where $p_n(w)$ is the probability distribution function
\begin{equation}
p_n(w) = \sum_{m k} p^B_m p^B_{n k|m} \delta(w-E^n_k+E_m)
\end{equation}
where the transition probability is $p^B_{n k|m} = \abs{\braket{E^n_k}{E_m}}^2$, $\ket{E_m}$ is the eigenstate of $H_B$ with eigenvalue $E_m$ and the environment populations are $p^B_m = \bra{E_m} \rho_B(0) \ket{E_m}$. The proof of Eq.~\eqref{prob} immediately follows from the definition of $p(w)$ by noting that the eigenstates of $H$ are the states $\ket{n}\otimes \ket{E^n_k}$. Physically, $p_n(w)$ represents the probability distribution function of the environment energy change in the quench $H_B\to H_B+B_n$, starting from the initial state $\rho_B(0)$. Since $\rho_B(0)$ is an equilibrium state, we obtain the Jarzynski equality~\cite{campisi11,Jarzynski97}
\begin{equation}\label{jarz}
\langle e^{-\beta w} \rangle_n = e^{-\beta \Delta F_B^n}
\end{equation}
where the average $\langle \cdots \rangle_n$ is calculated with respect to $p_n(w)$, and the free energy difference is $\Delta F_B^n = F^n_B - F_B$ with $F_B^n = -\ln(Z_B^n)/\beta$, $Z_B^n=\Tr{e^{-\beta(H_B+B_n)}}$ and $F_B=-\ln(Z_B)/\beta$.
Thus, from Eq.~\eqref{jarz} we get the Jarzynski equality
\begin{equation}\label{jarz1}
\langle e^{-\beta w}\rangle  = \sum_n p^S_n e^{-\beta \Delta F_B^n}= e^{-\beta \Delta F}
\end{equation}
where $\Delta F$ is the total free energy difference. In particular, in this form, the Eq.~\eqref{jarz1} relates the work done to the free energy differences of the environment.
From Eq.~\eqref{jarz1}, by using the Jensen's inequality we obtain the second law
\begin{equation}
\langle w \rangle \geq \Delta F =-\frac{1}{\beta}\ln\left(\sum_n p^S_n e^{-\beta \Delta F_B^n} \right)
\end{equation}
Since $f(x)=-\ln x$ is a convex function, we have
\begin{equation}
\Delta F \leq \sum_n p^S_n F_B^n - F_B\,.
\end{equation}
Conversely, by applying the Jensen's inequality to Eq.~\eqref{jarz}, we obtain
\begin{equation}
\langle w \rangle_n \geq F_B^n - F_B\,,
\end{equation}
from which, since $\langle w\rangle = \sum_n p^S_n \langle w\rangle_n$, we get the tighter lower bound of the work
\begin{equation}\label{bound}
\langle w \rangle \geq \sum_n p^S_n F_B^n - F_B \geq \Delta F\,.
\end{equation}
Of course, from Eq.~\eqref{bound} we also get a lower bound of the irreversible work $\langle w_{irr}\rangle = \langle w\rangle - \Delta F$.
We note that if all the operators $B_n$ commutate with the environment Hamiltonian $H_B$, i.e. $[H_B,B_n]=0$ for each $n$, the probability distribution function $p_n(w)$ simplifies into $p_n(w) = \sum_m p^B_m \delta(w-E^n_m+E_m)$.
Furthermore, we observe that our results inherent in the work probability distribution function and the lower bound can be immediately generalized to the case of time-dependent operators $B_n(t)$ by taking the transition probability $p^B_{nk|m}=\abs{\bra{E^n_k(t)} e^{-i H_B t} V_n(t)\ket{E_m}}^2$, where we have defined the unitary operator $V_n(t)= \mathcal T e^{-i\int_0^t B^I_n(s) ds}$, the time-ordering operator $\mathcal T$ and $B^I_n(t) = e^{i H_B t} B_n(t) e^{-i H_B t}$. Thus, in this case $\langle w \rangle_n$ reads
\begin{eqnarray}
\nonumber \langle w \rangle_n &=& \Tr{(V_n^\dagger(t) H_B V_n(t)-H_B)\rho_B(0)}\\
&& + \Tr{V_n^\dagger(t) B^I_n(t) V_n(t)\rho_B(0)}\,.
\end{eqnarray}
We recall that by considering the initial state $\rho(0)=\ket{\psi(0)}\bra{\psi(0)}\otimes \rho_B(0)$, the decoherence function takes the form $\Gamma_{nm}(t)=\ln\abs{\Tr{V_m^\dagger(t)V_n(t)\rho_B(0)}}$, then there seems to be no strong relation between the decoherence and the work done.
We note that by switching off the interaction at the final time the work will be equal to the environment energy change, such that we get as particular case the situation investigated in Ref.~\cite{Popovic21}.

\section{Physical examples}
We consider a system that is a qubit with Hamiltonian $H_S = -\omega \sigma_z/2$, where the matrix $\sigma_z$ is the third Pauli matrix and the parameter $\omega$ is the difference of energy levels of the system. We focus on the interaction $H_I=\sigma_z\otimes B$. By defining the eigenstates $\ket{0}$ and $\ket{1}$ of $\sigma_z$ with eigenvalues $1$ and $-1$, the interaction reads $H_I = \ket{0}\bra{0}\otimes B_{0} + \ket{1}\bra{1} \otimes B_{1}$ with $B_{0}=-B_{1} = B$. We have the work
\begin{equation}\label{qubit}
\langle w \rangle = \delta p^S \Tr{B \rho_B(0)}
\end{equation}
where we have defined the population unbalance $\delta p^S = p^S_0 - p^S_1=\tanh(\beta\omega/2)$, and the second moment
\begin{equation}\label{qubit2}
\langle w^2 \rangle = \Tr{B^2 \rho_B(0)}\,.
\end{equation}
We start by considering a bosonic environment with Hamiltonian $H_B=\sum_k \omega_k a_k^\dagger a_k$ and $B= \sum_k g_k (a_k + a_k^\dagger)$. We label with $k$ the modes of the environment having frequency $\omega_k$ and creation and destruction operators $a^\dagger_k$ and $a_k$, such that $[a_k,a_{k'}^\dagger]=\delta_{k k'}$ and the real parameters $g_k$ are the coupling constants.
We have that $\Tr{B \rho_B(0)}=0$, then $\langle w \rangle =0$ such that we need a non-linear interaction for obtaining $\langle w \rangle \neq 0$. Anyway, the higher work moments can be different from zero, for instance the second moment reads $\langle w^2 \rangle = \sum_k g^2_k \coth(\beta \omega_k /2)$.

We proceed by considering a fermionic environment with Hamiltonians $H_B= -t\sum_j (c_j^\dagger c_{j+1}+ c_{j+1}^\dagger c_j) - \mu \sum_j c_j^\dagger c_j$ and $B = \sum_j g_j c_j^\dagger c_j$. In particular, the parameter $t$ is the hopping amplitude, $\mu$ is the chemical potential and $c^\dagger_j$ and $c_j$ are creation and destruction operators, such that $\{ c_j, c_{j'}^\dagger\}=\delta_{j j'}$. In this case the work can be non-zero.
We focus on the particular case of a homogeneous coupling $g_j=g$, such that we have $B=g N$ where the number operator is $N=\sum_j c_j^\dagger c_j$. Since $[N,H_B]=0$, the work moments are related to the moments of the number operator by the equations $\langle w^{2n-1} \rangle = g^{2n-1} \delta p^S \Tr{N^{2n-1}\rho_B(0)}$ and $\langle w^{2n} \rangle = g^{2n} \Tr{N^{2n}\rho_B(0)}$.
Then, by  performing a Fourier transformation and expressing the environment Hamiltonian in the diagonal form $H_B=\sum \epsilon_k c_k^\dagger c_k$ with $\epsilon_k = -\mu -2t \cos k$, it is easy to calculate the work, which reads $\langle w \rangle = g\delta p^S\sum_k 1/(e^{\beta \epsilon_k}+1)$. In particular, the work can be positive or negative depending on the sign of the coupling $g$.
The lower bound of the work is $\sum_n p^S_n F^n_B - F_B = e^{\beta \omega/2} F_B(\mu-g)/Z_S+e^{-\beta \omega/2} F_B(\mu+g)/Z_S-F_B(\mu)$, where $Z_S=2\cosh(\beta \omega/2)$ and $F_B(\mu) = -\sum_k \ln(1+e^{-\beta \epsilon_k})/\beta$. If $j=1,\ldots,L$ and $g=g'/L$, in the limit $L\to \infty$ the bound is saturated, i.e. $\langle w\rangle=\sum_n p^S_n F^n_B - F_B = g' \delta p^S/\pi\int_{0}^\pi dk/(e^{\beta \epsilon_k}+1)$. We emphasize that we get the same bound for any time-dependent operator $B(t)$ such that at the final time is equal to $B=g N$.

\section{Conclusions}
In summary, we have investigated the quench thermodynamics of a pure decoherence process. In this case the work done on the system is equal to the heat exchanged with the environment. We have characterized the work probability distribution function, also with a fluctuation relation, a lower bound of the work done and by considering some physical examples.

We note that our results are different from the ones of Ref.~\cite{Popovic21} because we do not switch off the interaction at the final time. In particular, in the switching off process an amount of work is suddenly produced such that the total work is always non-negative, since the protocol becomes cyclic.

\appendix
\section{Strong coupling}
Since $H_S$ is constant, from Ref.~\cite{rivas20} we have the internal energy at the time $t$
\begin{equation}
\langle E_S(t)\rangle = \Tr{(H^*_S(\beta,t)+\beta \partial_\beta H^*_S(\beta,t) )\rho_S(t)}
\end{equation}
where $\rho_S(t) = \ParTr{B}{U_{t,0} \rho(0) U^\dagger_{t,0}}$ and
\begin{equation}
H^*_S(\beta,t) = -\frac{1}{\beta}\ln \left( \ParTr{B}{U_{t,0} e^{-\beta H_S}\otimes \rho_B(0)U^\dagger_{t,0}}\right)
\end{equation}
We consider the eigenvalue equation $H_S \ket{n} = \epsilon_n\ket{n}$ and the time evolved state $U_{t,0}\ket{n}\otimes\ket{E_k}=\ket{n}\otimes\ket{\phi^n_k(t)}$, then
\begin{eqnarray}
\nonumber U_{t,0}e^{-\beta H_S}\otimes \rho_B(0) U^\dagger_{t,0} &=& \sum_n e^{-\beta \epsilon_n} \ket{n}\bra{n}\otimes \sum_k \frac{e^{-\beta E_k}}{Z_B} \\
&& \times\ket{\phi^n_k(t)}\bra{\phi^n_k(t)}
\end{eqnarray}
Since $\Tr{\ket{\phi^n_k(t)}\bra{\phi^n_k(t)}}=1$, by taking the partial trace we get
\begin{equation}
\ParTr{B}{U_{t,0} e^{-\beta H_S}\otimes \rho_B(0)U^\dagger_{t,0}}=e^{-\beta H_S}
\end{equation}
then $H^*_S(\beta,t)=H_S$ and so the internal energy is $\langle E_S(t)\rangle = \Tr{U_{t,0}^\dagger H_S U_{t,0}\rho(0)}$.

\end{document}